\def\rfr#1{eq. (\ref{#1})}

\def\bar{\begin{eqnarray}}
\def\ear{\end{eqnarray}}
\def\bb{\bibitem}
\def\eqia{\begin{eqnarray}}
\def\eqfa{\end{eqnarray}}
\def\rp#1#2{{#1\over#2}}

\def\lb#1{\label{#1}}

\def\oc2{$\mathcal{O}(c^{-2})$}

\documentclass{aastex}
\usepackage{url}

\RequirePackage{color}

\begin{document}

\title{Will the recently approved LARES mission be able to measure the Lense-Thirring effect at $1\%$?}

\shorttitle{Will LARES be able to measure frame-dragging at $1\%$?}
\shortauthors{L. Iorio }

\author{Lorenzo Iorio }
\affil{INFN-Sezione di Pisa. Permanent address for correspondence: Viale Unit\`{a} di Italia 68, 70125, Bari (BA), Italy. E-mail: lorenzo.iorio@libero.it}

\begin{abstract}
After the approval by the Italian Space Agency of the LARES satellite, which should be launched at the end of 2009 with a VEGA rocket and whose claimed goal is a $\approx 1\%$ measurement of the general relativistic gravitomagnetic Lense-Thirring effect in the gravitational field of the spinning Earth, it is of the utmost importance to reliably assess the total realistic accuracy that can be reached  by such a mission. The observable is a linear combination of the nodes of the existing LAGEOS and LAGEOS II satellites and of LARES able to cancel out the impact of the first two even zonal harmonic coefficients of the multipolar expansion of the classical part of the terrestrial gravitational potential representing a major source of systematic error. While LAGEOS and LAGEOS II fly at altitudes of about 6,000 km, LARES should be placed at an altitude of 1,450 km. Thus, it will be sensitive to much more even zonals than LAGEOS and LAGEOS II. Their corrupting impact $\delta\mu$ has been evaluated by using the standard Kaula's approach up to degree $\ell=70$ along with the sigmas of the covariance matrices of eight different global gravity solutions (EIGEN-GRACE02S, EIGEN-CG03C, GGM02S, GGM03S, JEM01-RL03B, ITG-Grace02s, ITG-Grace03, EGM2008) obtained by five institutions (GFZ, CSR, JPL, IGG, NGA) with different techniques from long data sets of the dedicated GRACE mission.   It turns out $\delta\mu\approx 100-1,000\%$ of the Lense-Thirring effect. An improvement of $2-3$ orders of magnitude in the determination of the high degree even zonals would be required to constrain the bias to $\approx 1-10\%$.
\end{abstract}



 \section{Introduction}
In the weak-field and slow motion approximation, the Einstein field equations of general relativity resemble to the linear Maxwell equations of electromagntism. Thus, a gravitomagnetic field, induced by the off-diagonal components $g_{0i}, i=1,2,3$ of the space-time metric tensor related to the mass-energy currents of the source of the gravitational field, arises \citep{MashNOVA}. It affects in many ways the motion of test particles and electromagnetic waves \citep{Rug}. Perhaps, the most famous gravitomagnetic effects are the precession  of the axis of a gyroscope \citep{Pugh,Schi} and the Lense-Thirring\footnote{According to  \citet{Pfi07}, it would be more correct to speak about an Einstein-Thirring-Lense effect.} precessions \citep{LT} of the orbit of a test particle, both occurring in the field of a central slowly rotating mass like a planet.

The measurement of the gyroscope precession in the Earth's gravitational field has been the goal of the dedicated space-based\footnote{See on the WEB http://einstein.stanford.edu/} GP-B mission \citep{Eve,GPB} launched in 2004; its data analysis is still ongoing. The target accuracy was originally $1\%$, but it is now unclear if it will be possible to finally reach such a goal because of the onset of unexpected systematic errors during the mission.

In this paper we  critically discuss the possibility that the LARES mission, recently approved by the Italian Space Agency (ASI), will be able to measure the Lense-Thirring precession with an accuracy of the order of $1\%$. We recall that such a relativistic effect consists of a secular rate of the longitude of the ascending node $\Omega$ of the orbit of  a test particle
\begin{equation}\dot\Omega_{\rm LT} = \rp{2 G S}{c^2 a^3 (1-e^2)^{3/2}},\lb{let}\end{equation} where $G$ is the Newtonian constant of gravitation, $c$ is the speed of light in vacuum, $S$ is the proper angular momentum of the central body, $a$ and $e$ are the semimajor axis and the eccentricity, respectively, of the test particle's orbit.

In \citep{vpe76a,vpe76b}  it was proposed to measure the Lense-Thirring precession of the nodes $\Omega$ of a pair of counter-orbiting spacecraft to be launched in terrestrial polar orbits and endowed with drag-free apparatus to counteract the non-gravitational perturbations. A somewhat equivalent, cheaper version of such an idea was put forth in 1986 by \citep{Ciu86} who proposed to launch a passive, geodetic satellite in an orbit identical to that of the existing LAGEOS satellite apart from the orbital planes which should have been displaced by 180 deg apart. LAGEOS was put into orbit in 1976, followed by its twin LAGEOS II in 1992; they are passive spacecraft entirely covered by retroreflectors which allow for their accurate tracking through laser pulses sent from Earth-based ground stations according to the Satellite Laser Ranging (SLR) technique. They orbit at altitudes of about 6,000 km ($a_{\rm LAGEOS} =12,270$ km, $a_{\rm LAGEOS\ II}=12,163$ km) in nearly circular paths ($e_{\rm LAGEOS}=0.0045$, $e_{\rm LAGEOS\ II}=0.014$) inclined by 110 deg and 52.65 deg, respectively, to the Earth's equator. The Lense-Thirring effect for them amounts to about 30 milliarcseconds per year (mas yr$^{-1}$). The measurable quantity was, in the case of the proposal by \citet{Ciu86}, the sum of the nodes of LAGEOS and of the new spacecraft, later named LAGEOS III, LARES, WEBER-SAT, in order to cancel to a high level of accuracy the corrupting effect of the multipoles of the Newtonian part of the terrestrial gravitational potential which represent the major source of systematic error (see Section \ref{due}). Although extensively studied by various groups \citep{CSR,LARES}, such an idea was not implemented for many years.
In \citep{Ioretal02} it was proposed to include also the data from LAGEOS II by using a different observable.  Such an approach was proven in \citep{IorNA} to be potentially useful in making the constraints on the orbital configuration of the new SLR satellite less stringent than it was originally required in view of the recent improvements in our knowledge of the classical part of the terrestrial gravitational potential due to the dedicated CHAMP (http://www.gfz-potsdam.de/pb1/op/champ/index$\_$CHAMP.html) and, especially, GRACE (http://www.gfz-potsdam.de/pb1/op/grace/index$\_$GRACE.html, http://www.csr.utexas.edu/grace/) missions.

Reaching high altitudes and minimizing the unavoidable orbital injection errors is expensive; thus, it was explored the possibility of discarding LAGEOS and LAGEOS II using a low-altitude, nearly polar orbit for LARES \citep{LucPao01,Ciu06b}, but in \citep{Ior02,Ior07c} it was proven that such alternative approaches are not feasible. It was also suggested that LARES would be able to probe alternative theories of gravity \citep{Ciu04b}, but also in this case it turned out to be impossible \citep{IorJCAP,Ior07d}.

ASI recently made the following official announcement  (http://www.asi.it/SiteEN/MotorSearchFullText.aspx?keyw=LARES): ``On February 8, the ASI board approved funding for the LARES mission, that will be launched with VEGA's maiden flight before the end of 2008. LARES is a passive satellite with laser mirrors, and will be used to measure the Lense-Thirring effect.''  The italian version of the announcement yields some more information specifying that LARES, designed in collaboration with National Institute of Nuclear Physics (INFN), is currently under construction by Carlo Gavazzi Space SpA; its Principal Investigator (PI) is I. Ciufolini and its scientific goal is to measure at a $1\%$ level the  Lense-Thirring effect in the gravitational field of the Earth.
Concerning the orbital configuration of LARES, the ASI website says about VEGA that (http://www.asi.it/SiteEN/ContentSite.aspx?Area=Accesso+allo+spazio):
``[...] VEGA can place a 15.000 kg satellite on a low polar orbit, 700 km from the Earth. By lowering the orbit inclination it can launch heavier payloads, whereas diminishing the payload mass it can achieve higher orbits. [...]''    In the latest communication to INFN, Rome, 30 January 2008, \cite{INFN} writes that LARES will be launched with a semimajor axis of approximately 7600 km and an inclination between 60 and 80 deg.
More precise information can be retrieved in Section 5.1, pag 9 of the document Educational Payload on the Vega Maiden Flight
Call For CubeSat Proposals, European Space Agency,
Issue 1
11 February 2008, downloadable at
http://esamultimedia.esa.int/docs/LEX-EC/CubeSat$\%$20CFP$\%$20issue$\%$201.pdf.
    It is  written there that LARES will be launched into a circular orbit with altitude $h=1,200$ km, corresponding to a semimajor axis $a_{\rm LARES}=7,578$ km, and inclination $i=71$ deg to the Earth's equator.
    Recent informations\footnote{See on the WEB: http://www.esa.int/esapub/bulletin/bulletin135/bul135f$\_$bianchi.pdf.} point towards a launch in November 2009  with an altitude of\footnote{D. Barbagallo, ESA-ESRIN, personal communication to the author, September 2008.} $h=1,450$ km, i.e. $a=7,828$ km.
Such an orbital configuration will make difficult to obtain a total accuracy of the order of $1\%$, as we will show in Section \ref{due}.
\section{The major source of systematic error: the multipolar expansion of the Earth's gravitational potential}\lb{due}
In a realistic scenario the path of a probe is not only affected by the gravitomagentic field but also by a huge number of other competing orbital perturbations of gravitational and non-gravitational origin. The most important non-conservative accelerations  \citep{Mil87} are the direct solar radiation pressure, the Earth's albedo and various subtle thermal effects depending on the the physical properties of the satellite's surface and its rotational state \citep{Inv94,Ves99,Luc01,Luc02,Luc03,Luc04,Lucetal04,Ries03a}; however, the nodes  of LAGEOS-like satellites are directly sensitive to them at a $\approx 1\%$ level only. Much more important is the impact that the oblateness of the Earth has on the satellite's dynamics.
Indeed, the most insidious perturbations are those induced by the static part of the Newtonian component of the multipolar expansion in spherical harmonics\footnote{The relation among the  even zonals $J_{\ell}$ and the  normalized gravity coefficients $\overline{C}_{\ell 0}$ which are customarily determined in the Earth's gravity models, is $J_{\ell}=-\sqrt{2\ell + 1}\ \overline{C}_{\ell 0}$.} $J_{\ell}, \ell = 2,4,6,...$ of the gravitational potential of the central rotating mass \citep{Kau}: indeed, they affect the node with effects having the same signature of the relativistic signal of interest, i.e. linear trends which are orders of magnitude larger and cannot be removed from the time series of data without affecting the Lense-Thirring pattern itself as well. The only thing that can be done is to model such a corrupting effect as most accurately as possible and assessing, reliably and realistically, the impact of the residual mismodelling on the measurement of the frame-dragging effect.
The secular precessions induced by the even zonals of the geopotential can be written as
\begin{equation}\dot\Omega^{\rm geopot}=\sum_{\ell  =2}\dot\Omega_{.\ell}J_{\ell},\end{equation}
where the coefficients $\dot\Omega_{.\ell}, \ell=2,4,6,...$ depend on the parameters of the Earth ($GM$ and the equatorial radius $R$) and on the semimajor axis $a$, the inclination $i$ and the eccentricity $e$ of the satellite. For example, for $\ell=2$
we have
\begin{equation}\dot\Omega_{.2}=-\rp{3}{2}n\left(\rp{R}{a}\right)^2\rp{\cos i}{(1-e^2)^2};\end{equation}
 $n=\sqrt{GM/a^3}$ is the Keplerian mean motion.
They have been analytically computed up to $\ell=20$  in, e.g., \citep{Ior03}.
Their mismodelling can be written as
\begin{equation}\delta\dot\Omega^{\rm geopot}\leq \sum_{\ell  =2}\left|\dot\Omega_{.\ell}\right|\delta J_{\ell},\lb{mimo}\end{equation}
where $\delta J_{\ell}$ represents our uncertainty in the knowledge of the even zonal $J_{\ell}$ of degree $\ell$; a way to assess them consists of taking the sigmas of the covariance matrix of a global Earth gravity solution provided that they have been realistically calibrated. It turns out that using the node of only one satellite is not possible because the gravitomagnetic signal would be swamped by the much larger mismodelled classical precessions. Thus, it is necessary to use other satellites to enhance the signal-to-noise ratio.
The combination which will be used for measuring the Lense-Thirring effect with LAGEOS, LAGEOS II and LARES was obtained by  \citet{IorNA} according to a strategy put forth by \citet{Ciu96}; it is
\begin{equation} \dot\Omega^{\rm LAGEOS}+c_1\dot\Omega^{\rm LAGEOS\ II}+ c_2\dot\Omega^{\rm LARES},\ c_1=0.3586,\ c_2=0.0751.\lb{combi}\end{equation}
and, by construction, is not affected by $\delta J_2$ and $\delta J_4$.
The total Lense-Thirring effect, according to \rfr{let} and \rfr{combi}, amounts to 50.8 mas yr$^{-1}$.
The systematic percent error due to the mismodelling in the uncancelled even zonals $J_6, J_8,...$ can be conservatively evaluated as
\begin{equation}\delta\mu\leq \left(\rp{\sum_{\ell = 6}\left|\dot\Omega^{\rm LAGEOS}_{.\ell}+c_1\dot\Omega^{\rm LAGEOS\ II}_{.\ell}+ c_2\dot\Omega^{\rm LARES}_{.\ell}\right|\delta J_{\ell}}{50.8\ {\rm mas\ yr}^{-1}}\right)100\lb{bias}.\end{equation}
\section{The impact of the high degree even zonal harmonics on a low-orbit LARES}
Nowadays, contrary to even a few years ago, a growing number of global Earth's gravity solutions, produced by different institutions by analyzing longer and longer data sets from GRACE with different approaches and techniques are becoming available. Thus, it is possible to obtain more robust and reliable evaluations of the total accuracy which could be realistically obtained with LARES according to \rfr{combi}. The models we will use are EIGEN-GRACE02S \citep{eigengrace02s} and EIGEN-CG03C \citep{eigencg03c} from GFZ (Potsdam, Germany), GGM02S \citep{ggm02} and GGM03S \citep{ggm03} from CSR (Austin, Texas), ITG-Grace02s \citep{ITG} and ITG-Grace03
\citep{itggrace03s} from IGG (Bonn, Germany), JEM01-RL03B from JPL (NASA, USA) and EGM2008 \citep{egm2008} from National Geospatial-Intelligence Agency (NGA, USA).

Another very important point is that the low altitude of LARES  forces us to consider much more even zonals  with respect to the simpler original scenario for which a calculation up to degree $\ell=20$ would have been more than adequate. Instead, in all the previous evaluations, the calculations were truncated at $\ell=20$ \citep{IorNA} or even less \citep{Ciu06b}. Thus, we numerically computed the node precession coefficients $\dot\Omega_{.\ell}$ of LAGEOS, LAGEOS II and LARES up to $\ell=70$ with two different softwares according to the standard and widely used approach by \citet{Kau}. The figures determined in this way have been successfully calibrated by comparing them with those obtained from the coefficients analytically worked out by \citet{Ior03} up to $\ell=20$; the same values have been obtained.
\begin{table*}
\small
\caption{\label{tavola} Systematic percent error $\delta\mu$ $(\%)$ according to \rfr{bias} up to $\ell=70$. Second column: linear sum of the absolute values of the individual biased terms (SAV). Third column: square root of the sum of the squares of the individual biased terms (Root-Sum-Square, RSS). The orbital parameters of LARES are $a=7,828$ km, $i=71$ deg, $e=0$. For $\delta J_{\ell},\ \ell=2,4,...70$ the sigmas  of the covariance matrices of EIGEN-GRACE02S (calibrated), EIGEN-CG03C (calibrated), GGM02S (formal), GGM03S (calibrated), ITG-Grace02s (formal), ITG-Grace03 (formal), JEM01-RL03B (formal) and EGM2008 (calibrated) have been used in both cases. }

\begin{tabular}{@{}ccc@{}}
\hline
 Model  & $\delta\mu$ (up to $\ell=70$: SAV) & $\delta\mu$ (up to $\ell=70$: RSS)\\
\tableline
EIGEN-GRACE02S  & $4,443\%$ &  $2,563\%$\\
EIGEN-CG03C  &  $2,177\%$ &  $1,256\%$\\
GGM02S  & $3,595\%$ &  $2,062\%$\\
GGM03S  & $911\%$ &  $524\%$\\
ITG-Grace02s  & $63\%$ &  $36\%$\\
ITG-Grace03   & $58\%$ &  $33\%$\\
JEM01-RL03B  & $235\%$ &  $135\%$\\
EGM2008  & $1,307\%$ &  $754\%$\\
\hline
 \end{tabular}
\end{table*}
As can be noted from an inspection of Table \ref{tavola}, obtained by using the covariance sigmas for the aforementioned Earth's gravity models, the fact that LARES will be much more sensitive to the high degree even zonal harmonics than LAGEOS and LAGEOS II  introduces a corrupting effect on the measurement of the Lense-Thirring precession which makes the expectation for a $\approx 1\%$ accuracy difficult to be implemented; it is true also in the cases in which only the formal, statistical sigmas were available. In Table \ref{tavola} we quote both the upper bound obtained by linearly adding the absolute values of the individual mismodelled terms (SAV) and a more optimistic evaluation obtained by computing the square root of the sum of the squares of the individual biased terms (Root-Sum-Square, RSS); both the approaches yield very large figures.
It turns out that improvements of at least $3-2$ orders of magnitude in the even zonals, likely unattainable in the foreseeable future, would be needed to constrain $\delta\mu$ at the $\approx 1-10\%$ level. It would be desirable that other researchers will independently repeat the calculations with different computational approaches.

%
%
%
%

%
%
\section{Conclusions}
In view of the recent approval of the LARES mission, to be launched at the end of 2009 with a VEGA rocket into a circular orbit with semimajor axis $a=7,828$ km and inclination $i=71$ deg, we discussed the possibility that the LAGEOS, LAGEOS II and LARES satellites will be able to measure the Lense-Thirring effect at an accuracy of the order of $1\%$. The observable will be a linear combination of the nodes of the three spacecraft which, in fact, should be affected by the non-conservative perturbations at a $\approx 1\%$ level. A major source of systematic bias is represented by the uncertainty in the static part of the even zonal harmonic coefficients of the multipolar expansion of the Newtonian component of the terrestrial gravitational potential. The low altitude of LARES, 1,450 km with respect to about 6,000 km of LAGEOS and LAGEOS II, will make its node sensitive to much more even zonals than its two already orbiting twins; it turns out that, by using the sigmas of the covariance matrices of some of the latest global Earth's gravity solutions based on long data sets of the dedicated GRACE mission, the systematic bias due to the mismodelled even zonal harmonics up to $\ell=70$ will amount to $\approx 100-1,000\%$.  An improvement of $3-2$ orders of magnitude in the determination of the high degree even zonals would be required to reduce the systematic error to $\approx 1-10\%$ level.

\section*{Acknowledgments}
 I am grateful to J Ries (CSR) and M Watkins (JPL) for having provided me with the spherical harmonics' coefficients of the GGM03S and JEM01-RL03B models along with their sigmas.  I thank D Barbagallo (ESA-ESRIN) for the latest orbital parameters of LARES.


 \end{document}